\documentclass[%
 reprint,
 amsmath,amssymb,
 aps,
]{revtex4-1}

\usepackage{graphicx}
\usepackage{dcolumn}
\usepackage{natbib}
\usepackage{filecontents}
\usepackage{bm}

\begin{document}

\preprint{APS/123-QED}

\title{Unified mechanism of the surface Fermi level pinning in III-As nanowires }

\author{P. A. Alekseev}
\email{npoxep@gmail.com}
 \affiliation{Ioffe Institute, Saint-Petersburg 194021, Russia}
\author{M. S. Dunaevskiy}%
\affiliation{Ioffe Institute, Saint-Petersburg 194021, Russia}
\affiliation{ITMO University, Saint-Petersburg 197101, Russia}

\author{G. E. Cirlin}
 \affiliation{ITMO University, Saint-Petersburg 197101, Russia}
\affiliation{Saint-Petersburg Academic University, Saint-Petersburg 194021, Russia}
\affiliation{Institute of Analytical Instrumentation, Saint-Petersburg 198095, Russia}
\author{R. R. Reznik}
\affiliation{ITMO University, Saint-Petersburg 197101, Russia}
\affiliation{Saint-Petersburg Academic University, Saint-Petersburg 194021, Russia}
\affiliation{Institute of Analytical Instrumentation, Saint-Petersburg 198095, Russia}
\author{A. N. Smirnov}
\affiliation{Ioffe Institute, Saint-Petersburg 194021, Russia}
\author{V. Yu. Davydov}
\affiliation{Ioffe Institute, Saint-Petersburg 194021, Russia}
\author{V. L. Berkovits}
\affiliation{Ioffe Institute, Saint-Petersburg 194021, Russia}

\date{\today}

\begin{abstract}
Fermi level pinning at the oxidized (110) surfaces of III-As nanowires (GaAs, InAs, InGaAs, AlGaAs) is studied. Using scanning gradient Kelvin probe microscopy, we show that the Fermi level at oxidized cleavage surfaces of ternary Al$_{x}$Ga$_{1-x}$As (0$\le$x$\le$0.45) and Ga$_{x}$In$_{1-x}$As (0$\le$x$\le$1) alloys is pinned at the same position of 4.8$\pm$0.1 eV with regard to the vacuum level. The finding implies a unified mechanism of the Fermi level pinning for such surfaces. Further investigation, performed by Raman scattering and photoluminescence spectroscopy, shows that photooxidation of the Al$_{x}$Ga$_{1-x}$As and Ga$_{x}$In$_{1-x}$As nanowires leads to the accumulation of an excess arsenic on their crystal surfaces which is accompanied by a strong decrease of the band-edge photoluminescence intensity. We conclude that the surface excess arsenic in crystalline or amorphous forms is responsible for the Fermi level pinning at oxidized (110) surfaces of III-As nanowires.
\begin{description}
\item[PACS numbers]
73.20.At. 
\end{description}
\end{abstract}

\maketitle
\section{\label{sec:level1}Introduction}
Compound III-As semiconductors (where III atoms are Ga, In, Al) and their ternary alloys, owing to a direct band structure and high carrier mobility, are prospective materials for high electron mobility transistors (HEMT) \cite{mimura1980new}, as well as for a large number of optoelectronic devices\cite{faist1998high,huffaker19981}. In the last decade, following the general trend to reduce the device sizes down to the nanometer scale, III-As nanowires (NWs) began to be employed for device engineering \cite{bryllert2006vertical,saxena2013optically,colombo2009gallium}. In addition to the advantageous essential properties of III-As materials, these nanostructures have their own beneficial features which could be used in nano- and optoelectronics. In particular, the III-As nanowires can be grown on silicon substrates, which facilitates their compatibility with silicon-based device structures \cite{gudiksen2002growth,maartensson2004epitaxial}. Also, these nanowires possess an increased light absorption and can be employed, therefore, in the fabrication of solar cells with increased efficiency \cite{krogstrup2013single}. However, conventional oxidized surfaces of III-As compounds have a high density of surface states that is related, generally, to surface defects \cite{cowley1965surface}. These states are responsible for the surface Fermi level pinning, formation of the near- surface depletion layer, and surface recombination. Owing to the high surface/volume ratio, these surface-induced, negative effects, for III-As nanowires are anticipated to be drastically enhanced. For example, the Fermi level mid-gap pinning that occurs in low-doped GaAs and AlGaAs NWs can totally switch off the conductivity along the nanowire, as well as the photoluminescence \cite{chang2012electrical}. Fixation of the Fermi level in the conduction band known for InAs surfaces facilitates formation of the ohmic contact in n-type nanowires, but gives rise to inversion of the channel conductivity in p-type nanowires \cite{li2007remote}. It is clear that a comprehensive understanding of the pining induced phenomena is of fundamental importance and also necessary for the design of device structures based on III-As nanowires. 
Conventionally, III-As nanowire surfaces are formed by (110) facets. Active surfaces of semiconductor lasers have the same orientation. Fermi level pinning does not occur at clean, cleavage (110) faces of III-As bulk semiconductors, which corresponds to the so-called flat-band condition \cite{monch1986virtual}. However, even a submonolayer deposition of O, Ga, In, Au, Al, and other elements leads to fixation of the surface Fermi level at a position that differs from its position in a bulk crystal. The surface pinning positions are usually measured relative to the tops of valence or conduction bands. In such coordinates, the pinning positions for III-As semiconductors are independent of the nature of the adsorbed atoms \cite{spicer1979new}. As a consequence, the surface states responsible for the pinning are now assigned to intrinsic surface defects. The pinning positions for (110) GaAs are ~ 0.75 eV and 0.55 eV above the valence band for n- and p-type materials, respectively \cite{spicer1980unified}. For the (110) InAs surface, the Fermi level is fixed at 0.13 eV above the bottom of the conduction band for n- and p-type materials \cite{baier1986oxidation}. For Al$_{x}$Ga$_{1-x}$As ternary alloys, the Fermi level pinning has been studied much less. Note that despite the long period of investigations, the nature of the surface states responsible for the Fermi level pinning in III-arsenides is still a controversial issue.
We study the Fermi level pinning at oxidized surfaces of the III-As nanowires. We determine pinning positions on the cleavage surfaces of the III-As heterostructures and study the behavior of the photoluminescence for III-As nanowires under photooxidation. Based on the obtained results, a unified mechanism for the surface Fermi level pinning in the III-As structures and nanowires is proposed.

\section{\label{sec:level1}Results and discussion}

Experimentally, the pinning positions are determined by scanning gradient Kelvin probe microscopy \cite{dunaevskiy2011kelvin}. At each scanning point, this technique measures the surface potential, which is the difference between the work functions of a microscope probe and of a semiconductor surface. The latter value presents the energy gap between the vacuum level and the surface Fermi level. In the case of the cleavage faces of III-As heterostructures, such measurements provide a determination of the pinning position for each heterolayer \cite{usunami1998cross}. To verify the validity of such measurements, we measured the surface potential values on the cleavage face of a p-i-n GaAs structure. The measurements were performed on naturally oxidized surfaces under atmospheric conditions. Figure 1(a) shows the measured profile of the surface potential value along the cleaved n$^{+}$ (5$\times$10$^{18}$ cm$^{-3}$)-n (1$\times$10$^{15}$)-p$^{+}$ (5$\times$10$^{18}$) layers of the structure. From Fig. 1(a), one can conclude that the energy gap between pinning positions for the p$^{+}$ and n$^{+}$ layers was approximately 0.2 eV, which was in agreement with reference \cite{spicer1980unified}. For the weakly doped intermediate layer, the pinning position occurred approximately in the middle between those found for the p$^{+}$ layer and the n$^{+}$ substrate.
To study the Fermi level pinning at the (110) surface of ternary Al$_{x}$Ga$_{1-x}$As, we used the cleavage face of a laser-diode heterostructure Al$_{x}$Ga$_{1-x}$As \cite{vinokurov2005high}. The heterostructure was grown on n$^{+}$ GaAs substrate and consisted of the following layers: $^{+}$-emitter Al$_{0.4}$Ga$_{0.6}$As/ p-waveguide Al$_{0.3}$Ga$_{0.7}$As/ p-emitter Al$_{0.4}$Ga$_{0.6}$As/, and contact p$^{+}$-GaAs layer. The profile of the surface potential measured across the heterostructure cleavage face is shown in Fig. 1(b). As was observed in Fig. 1(a), the difference between the pinning positions for p$^{+}$ and n$^{+}$ GaAs layers was ~ 0.2 eV. 
As seen from the Fig. 1(b), the pinning positions for the heavily doped n$^{+}$ Al$_{0.4}$Ga$_{0.6}$As layer and for the heavily doped n$^{+}$  GaAs substrate were very close. For weakly doped p-Al$_{x}$Ga$_{1-x}$As layers, the Fermi level was pinned approximately in the middle between the positions for the n$^{+}$ and p$^{+}$ layers.
\begin{figure}[h]
\center{\includegraphics{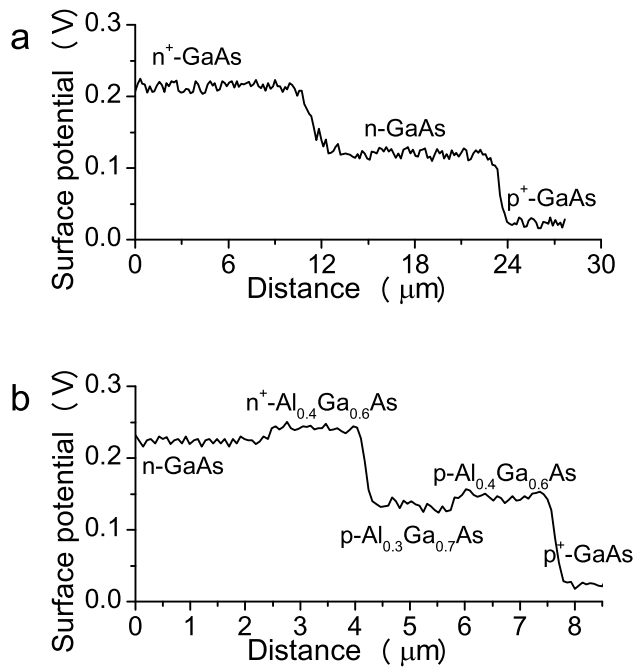}}
\caption{Profiles of the surface potential measured across cleavage of (a) p-i-n GaAs structure, (b) GaAs/AlGaAs laser diode heterostructure}
\label{FIG.}
\end{figure}

Approximately 20 heterostructures of Al$_{x}$Ga$_{1-x}$As, in which the Al concentration x varied in the range of 0.1-0.4, were studied. For all x values, the gap between the pinning positions for doped n$^{+}$ and p$^{+}$ layers did not exceed 0.25 eV, while for the weakly doped layers, the pinning positions occurred near the center of this gap. The pinning positions determined with regard to the vacuum level are plotted in Fig. 2(a). As can be seen, for the (110) surface of Al$_{x}$Ga$_{1-x}$As (0$\le$x$\le$0.45), the Fermi level was fixed near 4.8$\pm$0.1 eV and near 4.9$\pm$0.1 eV for the n-and p-type materials, respectively.

To plot the analogous dependence for oxidized surfaces of n-Ga$_{x}$In$_{1-x}$As (0$\le$x$\le$1), we analyzed the data published in Refs. \cite{spicer1980unified,baier1986oxidation,mullin1983surface}. In Fig. 2(b), the blue dots mark the known surface pinning positions for n-InAs (Ref. \cite{baier1986oxidation}), n-Ga$_{0.47}$In$_{0.53}$As (Ref. \cite{mullin1983surface}), and n-GaAs (Ref. \cite{spicer1980unified}). All these points occurred again at the same distance of 4.8$\pm$0.1 eV from the vacuum level. A similar result was obtained by Weider who observed that the pinning positions for the components of the Ga$_{x}$In$_{1-x}$As metal - insulator-semiconductor (MIS) structures for all x were also equidistant from the vacuum level \cite{wieder2003surface}. Thus, the obtained results unambiguously demonstrated that at the oxidized (110) surface of III-As compounds, the Fermi levels were fixed at the same energy position calculated from the vacuum level. One can also observe from Fig. 2(b) that for x$\approx$0.2, the positions of the Fermi level coincided with the bottom of the conduction band, which corresponded to the so-called flat band condition. This was in agreement with the recent studies on n-Ga$_{x}$In$_{1-x}$As nanowires\cite{speckbacher2016direct}. 

\begin{figure}[h]
\center{\includegraphics{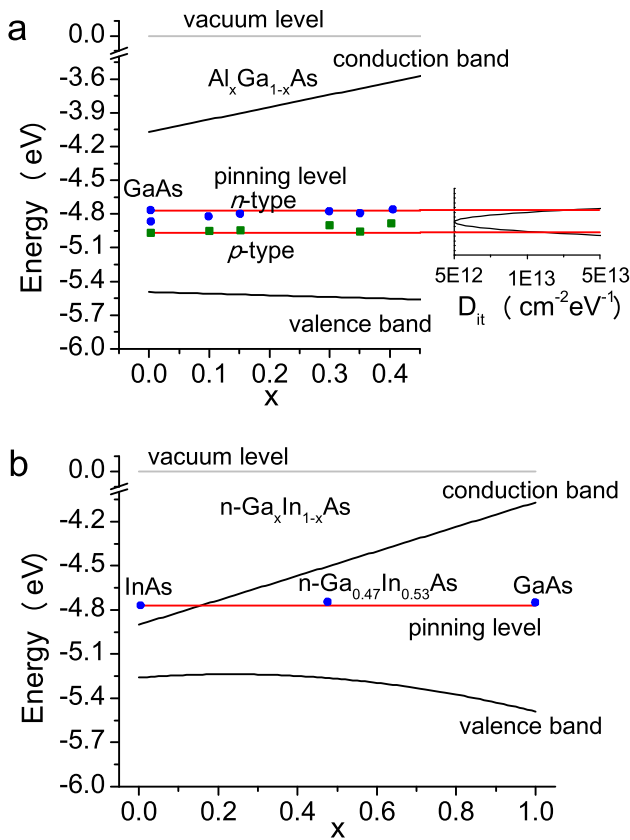}}
\caption{Band diagrams for III-As ternary alloys, red lines indicate positions of the surface Fermi level pinning. (a) - Fermi level positions of Al$_{x}$Ga$_{1-x}$As heterolayers measured by Kelvin probe microscopy. Blue dots indicate positions for n-type heterolayers, green squares - for p-type. The “U”-shape curve presents a distribution of the surface states density versus the same ordinate axis; (b) Fermi level positions for n-Ga$_{x}$In$_{1-x}$As surfaces taken from ref. \cite{spicer1980unified,baier1986oxidation,mullin1983surface}}
\label{FIG.}
\end{figure}

Assuming that the distribution of the surface states density had a "U" shape shape\cite{hasegawa2010surface}, we concluded that the minimum of this distribution was at a distance of ~ 4.9 eV from the vacuum level (see Fig. 2(a)), and the width of the distribution did not exceed 0.25 eV. The density of the surface states on the oxidized surface of III-As semiconductors in the distribution minimum has been reported to exceed 10$^{12}$ cm$^{-2}$eV$^{-1}$ (ref. \cite{hasegawa2010surface}). Therefore, for low doped p- and n- materials, the Fermi level should be fixed in the vicinity of the distribution minimum. With an increase of the doping level, the surface states at the distribution minimum began to be filled and the Fermi level moved along the distribution curve. As a result, the Fermi level moved upward or downward in the n- and p-type materials, respectively.

The higher the doping level, the larger was the shift of the Fermi level pinning position from the distribution minimum. However, owing to the sharp increase of the surface state density at a distance of ~ 0.1 eV from the distribution minimum, a further increase of the doping level did not shift the pinning position. It should also be noted that the value of the surface states density at the distribution minimum for an oxidized surface can vary from 10$^{12}$ up to 10$^{13}$ cm$^{-2}$eV$^{-1}$ which can shift the “U” curve in Fig. 2(a) as a whole to the lower or higher state density side of the plot.
The presented data unambiguously indicates the existence of a unified mechanism of the Fermi level pinning at oxidized surfaces of III-As semiconductor structures. Indeed, the positions of the surface Fermi level relative to the vacuum level were the same for all investigated III-As compounds and their ternary alloys. We suppose that the Fermi level pinning is caused by atoms of excess arsenic emerging at the III-As semiconductor surfaces under oxidation. Bearing this idea in mind, we studied the photooxidation of GaAs, Ga$_{x}$In$_{1-x}$As, and Al$_{x}$Ga$_{1-x}$As nanowires in which, as has already been mentioned, the pinning-induced effects were strongly enhanced. 
It is known that in air ambient, an intense optical irradiation accelerates the surface oxidation of GaAs and InAs nanowires \cite{yazji2011local,tanta2016morphology}. A photooxidation-induced decrease of the band-gap photoluminescence (PL) intensity for GaAs nanowires is observed \cite{yazji2011local}. We have recently shown that photooxidation leads to the formation of a double layer c-As/ GaO$_{x}$ at the GaAs nanowire surface and it is the c-As layer that contacts the crystal lattice of the nanowire \cite{alekseev2017observing}.

The photooxidation of III-As NWs was studied using a combination of Raman scattering and photoluminescence (PL) spectroscopy. The former technique was to control chemical composition of the NW surfaces. The experiments were carried out at room temperature (300 K) using Horiba Jobin Yvon T64000 and LabRAM HR spectrometers equipped with a Linkam THMS600 temperature-controlled microscope stage. The measurements were performed with continuous-wave (cw) excitation using the 532 nm laser line of a Nd: YAG laser. We used a Mitutoyo 100× NIR (NA=0.90) long working-distance objective lens to focus the incident beam into a spot of ~1 1 $\mu$m diameter, which was sufficient to measure the PL signal from a separate nanowire transferred to a Si/ SiO$_{2}$ substrate.

\begin{figure*}[!tbp]
\center{\includegraphics{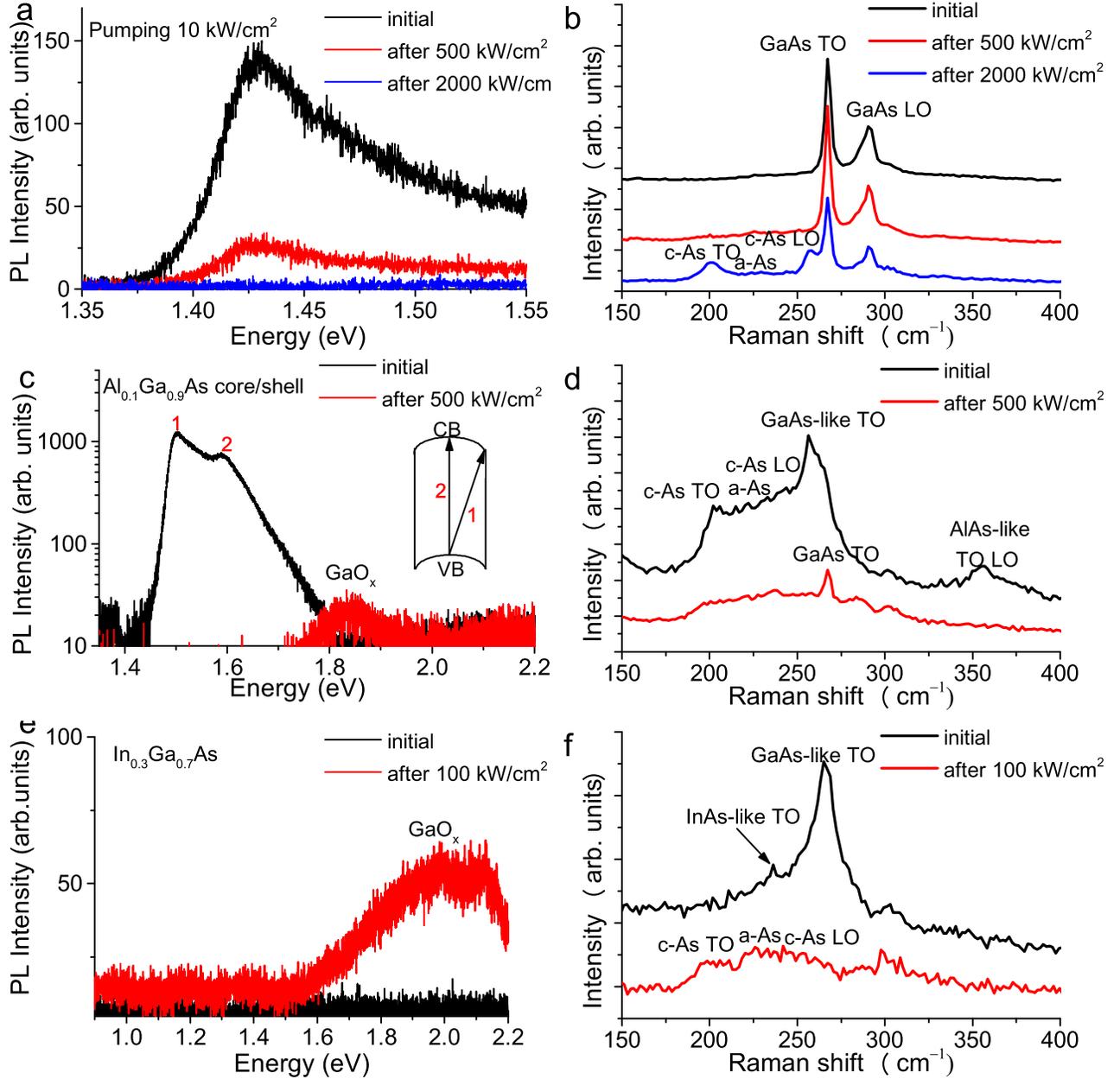}}
\caption{Photoluminescence (a), (c), (e) and Raman (b), (d), (f) spectra measured on single NWs before and after intensive laser irradiation. (a-b) for GaAs NW, (c-d) for Al$_{0.1}$Ga$_{0.9}$A core/shell NW, (e-f) for Ga$_{0.7}$In$_{0.3}$As NW}
\label{FIG.}
\end{figure*}

The studied GaAs, Ga$_{0.7}$In$_{0.3}$As, and Al$_{0.1}$Ga$_{0.9}$As nanowires were grown on Si substrates by the molecular beam epitaxy (MBE) technique using the vapor-liquid solid growth procedure. Gold droplets were used as catalyst. The Al$_{x}$Ga$_{1-x}$As nanowires had the core/shell structure Al$_{0.1}$Ga$_{0.9}$As/Al$_{0.2}$Ga$_{0.8}$As\cite{dubrovskii2016origin}, in which the first component was the core and the second component was the shell. Scanning electron microscopy images of the NWs are presented in Fig. S1 in the Supplemental material. The experimental procedure was as follows. First, photoluminescence and Raman scattering spectra were measured for the as-received nanowires. Then, the nanowires were exposed to successive one-minute laser irradiations with a power density of 100, 500, and 2000 kW/cm$^{2}$. After each laser-light exposure, the Raman scattering and PL spectra were measured at a low pumping density of 10 kW/cm$^{2}$. 
Figure 3(a) and 3(b) show the spectra obtained for a GaAs nanowire. It can be seen that after 500 kW/cm$^{2}$beam irradiation, the intensity of the initial PL spectrum (black curve in Fig. 3(a)) decreased by approximately a factor of 10. After the third irradiation (2000 kW/cm$^{2}$), the band-edge PL was hardly detected. In parallel, the intensive laser irradiation resulted in the appearance and growth of the additional lines in the Raman scattering spectra in the spectral region of 200-260 cm$^{-1}$ assigned to TO and LO phonons in amorphous and crystalline arsenic \cite{yazji2011local} (see Fig. 3(b)). These results clearly provide evidence that the surface excess arsenic in the crystalline or amorphous forms is the origin of the surface states responsible for surface nonradiative recombination and for the surface Fermi level pinning. These observations agree with the already known results observed on the photooxidation of GaAs nanowires \cite{yazji2011local,alekseev2017observing}.

Figure 3(c) and 3(d) show the optical spectra obtained for a Al$_{0.1}$Ga$_{0.9}$As/Al$_{0.2}$Ga$_{0.8}$As core/shell nanowire. The initial PL spectrum in Fig. 3(c) revealed a doublet whose components at 1.5 and 1.6 eV were assigned to the band-edge PL of the nanowire core (1.6 eV) and to the core-shell transitions (1.5 eV). The initial Raman scattering spectrum, excluding the lines of GaAs-like and AlAs-like TO phonons \cite{jusserand1981raman}, also exhibited the lines corresponding to the phonons in c- and a-arsenic (see Fig. 3(d)). This arsenic appeared owing to the greater thickness of the natural oxide layer in AlGaAs than that in GaAs \cite{ankudinov1999nanorelief}. However, the PL intensity of AlGaAs nanowires significantly exceeded the PL of GaAs nanowires, apparently because the NW shell with a higher aluminum content served as a passivation coating for the NW core. Since the NW shell was not fully oxidized, the oxide did not reach the core/shell interface. After irradiation of the nanowire by a beam with the power density of 500 kW/cm$^{2}$, the band-edge PL was not detected, which indicated that the PL intensity decreased by at least 3 orders of magnitude. At the same time, a weak PL appeared near 1.8 eV, which was apparently related to the formation of the GaO$_{x}$ layer\cite{alekseev2017observing}. The observed elimination of the band-edge PL intensity arose from the entire oxidation of the core / shell interface. In the Raman scattering spectrum, a weak line associated with a GaAs-like TO phonon remained. It should be noted that after laser irradiation, the broad arsenic-related spectral pedestal remained as well.

Owing to a small diameter and low doping level, the Ga$_{0.7}$In$_{0.3}$As nanowire was completely depleted by charge carriers. Therefore, at the pumping density of 10 kW/cm$^{2}$, the band-edge PL in the nanowire was not observed (Fig. 3(e)). The increase of the pump density resulted in observation of the PL \cite{alekseev2014nitride}, but in ambient conditions, it also produced a complete oxidation of the nanowire. In the initial Raman spectrum shown in Fig. 3(f), InAs-like and GaAs-like TO phonons were visible \cite{hertenberger2012high}. After laser irradiation with a power density of 100 kW/cm$^{2}$, these lines disappeared and, instead, the lines corresponding to crystalline and amorphous arsenic appeared. The weak PL peak observed in Fig. 3(e) in the region 1.7-2.2 eV could be explained by the formation of GaO$_{x}$ /Ga$_{2}$O$_{3}$ oxides.

Thus, under photooxidation, the intensity of the band-edge photoluminescence in III-As nanowires decreased with the simultaneous growth of the arsenic layer only on the crystal nanowire surface (see Fig. S2 in the Supplemental material). In addition, for all III-As compounds, fixation of the Fermi level at the oxidized (110) surfaces occurred at the same energy interval from the vacuum level. We propose that the surface arsenic was responsible for the Fermi level pinning in III-As semiconductors. Indeed, it was shown in Ref. \cite{chiang1989arsenic} that deposition of As on the clean (110) surface led to pinning of the Fermi level. Annealing of the oxidized surface provided a sublimation of As and led to the unpinning of the Fermi level. Unlike the surface As, the Ga-oxide components of the oxide layers in III-As compounds did not produce the Fermi level pinning. In Ref. \cite{passlack1997low}, the Ga$_{2}$O$_{3}$ layer was deposited on the GaAs surface and, as a result, the density of mid-gap surface states decreased down to 10$^{10}$ cm$^{-2}$eV$^{-1}$ for the sharp interface between GaAs and Ga$_{2}$O$_{3}$.

We now discuss why the surface arsenic created the surface-gap states in III-As. To explain the Fermi level pinning in such compounds, the approaches of the intrinsic defects or of effective work function are conventionally used. In the former approach, initially suggested by Spicer, the electronic states of surface gallium and arsenic vacancies were considered to produce pinning of the surface Fermi level \cite{spicer1980unified}. Later, more detailed calculations indicated that the mid-gap levels could belong to As-antisite defects whose existence at the oxidized GaAs surface was also confirmed experimentally \cite{stesmans2013asga+}. Very recent studies of the GaAs/oxide interface using density functional theory (DFT) indicated that the mid-gap Fermi level pinning could be caused by more complicated defect complexes, which included an As vacancy (V$_{As}$) and two As antisites (As$_{Ga}$)\cite{colleoni2015fermi}. The latter defect also emerged in the form of a V$_{Ga}$-As$_{Ga}$ complex. As a criticism of the native defects approach, a photo-washing experiment is usually cited. In this experiment, the PL intensity of GaAs samples immersed in water was measured \cite{wilmsen1988characterization}. The PL intensity was observed to increase by an order of magnitude after dissolution of surface arsenic. Moreover, a decrease of the surface states density down to 10$^{11}$ cm$^{-2}$eV$^{-1}$ and the unpinning of the surface Fermi level were demonstrated \cite{offsey1986unpinned}. It is unlikely that an aqueous solution could remove surface native defects. Also, the removal of the surface arsenic layer from GaAs by plasma etching was found to produce a decrease of the surface states density \cite{callegari1989unpinned}.

The effective work function approach, which was presented by Woodal and Freeouf, assumed that surface chemical reactions induced by oxidation or deposition of metals on the III-As surfaces led to the formation of the surface arsenic layer \cite{woodall1981gaas}. If the effective work function $\phi$$_{eff}$ is assigned to this As layer, the level pinning position and Schottky barrier height for the n-type materials can be determined as $\phi$$_{bn}$ = $\phi$$_{eff}$-$\chi$, where $\chi$ is the electron affinity for the III-As semiconductor. There are not too much data on the work function of arsenic. It is known that $\phi$$_{eff}$ =4.66 eV and 5.1 eV for an amorphous \cite{taft1949photoelectric} and crystalline As layer \cite{middleton2002}, respectively. The theoretically predicted value of $\phi$$_{eff}$ has been reported as 4.9 eV \cite{michaelson1977work}. In our case, the pinning positions in Figure 2(a,b) is in a good agreement with the work function value on the level of 4.8-4.9 eV. However, it is difficult to explain the shift of the Fermi level pinning position by 0.2 eV observed in GaAs when passing from p$^{+}$ to n$^{+}$ materials (Fig. 1(a)) using the effective work function approach. 

However, the approach using surface native defects remains as partially applicable. Computations in the framework of numerical models indicated that arsenic defect complexes, such as As-As dimer/dangling bond, are good candidates for amphoteric mid-gap defect levels. Such models also describe the abrupt interfaces of GaAs/Al$_{2}$O$_{3}$ quite well \cite{robertson2015defect}. 
Apparently, the real structure of the natural oxide layer in III-As compounds is quite complex. In particular, surface arsenic can be of a few monolayers of thickness, as has been shown experimentally \cite{alekseev2017observing}. To properly explain the Fermi level pinning in III-As compounds and their ternary alloys, the models taking into account surface multi-particle and even cluster arsenic complexes are required.
\section{\label{sec:level1}Conclusions}
In conclusion, the oxidation or deposition of metal on (110) surfaces of III-As semiconductor nanowires provides the formation of surface excess arsenic. The formed arsenic creates surface states which pin the Fermi level at the same distances from the vacuum level of 4.8$\pm$0.1 eV and of 4.9$\pm$0.1 eV for n-type and p-type materials, respectively. These findings are vitally important for modelling and designing electronic and optoelectronic devices based on III-As nanowires. Based on these data, we can predict that the pinning position in ternary GaAs$_{x}$Sb$_{1-x}$ NWs for (0.01$\le$x$\le$0.5) is situated in the valence band, and explains the formation of an ohmic contact to GaAs$_{0.5}$Sb$_{0.5}$\cite{li2017near}.

\acknowledgments

The reported study was funded by RFBR according to the Research Project 16-32-60147 mol-a-dk. M.S.D. acknowledge for financial support the Government of Russian Federation (Grant 074-U01). The nanowire samples were grown under the support of Russian Science Foundation (Project No 14-12-00393). The authors thank S.O. Slipchenko for providing bulk heterostructures samples.

\bibliography{ref}

\end{document}